\documentclass[12pt]{article}
\pdfoutput=1

\usepackage{amsfonts,amsthm,amsmath,amssymb,graphicx,hyperref,color,yfonts}

\newcommand{\bea}{\begin{eqnarray}\displaystyle}
\newcommand{\eea}{\end{eqnarray}}

\begin{document}
\makeatletter
\@addtoreset{equation}{section}
\makeatother
\renewcommand{\theequation}{\thesection.\arabic{equation}}
\vspace{1.8truecm}

{\LARGE{ \centerline{\bf Holography of Information in AdS/CFT}}}  

\vskip.5cm 

\thispagestyle{empty} 
\centerline{{\large\bf Robert de Mello Koch$^{a,}$\footnote{{\tt robert@zjhu.edu.cn}} and
Garreth Kemp$^{b,}$\footnote{garry@kemp.za.org}}}
\vspace{.8cm}
\centerline{{\it $^{a}$School of Science, Huzhou University, Huzhou 313000, China,}}
\vspace{.8cm}
\centerline{{\it $^{a}$School of Physics and Mandelstam Institute for Theoretical Physics,}}
\centerline{{\it University of the Witwatersrand, Wits, 2050, }}
\centerline{{\it South Africa }}
\vspace{.8cm}
\centerline{{\it $^{b}$Department of Mathematics and Applied Mathematics,}}
\centerline{{\it University of Johannesburg, Auckland Park, 2006, South Africa.}}

\vspace{1truecm}

\thispagestyle{empty}

\centerline{\bf ABSTRACT}

\vskip.2cm 
The principle of the holography of information states that in a theory of quantum gravity a copy of all the information available on a Cauchy slice is also available near the boundary of the Cauchy slice. This redundancy in the theory is already present at low energy. In the context of the AdS/CFT correspondence, this principle can be translated into a statement about the dual conformal field theory. We carry out this translation and demonstrate that the principle of the holography of information holds in bilocal holography.

\setcounter{page}{0}
\setcounter{tocdepth}{2}
\newpage
\tableofcontents
\setcounter{footnote}{0}
\linespread{1.1}
\parskip 4pt

{}~
{}~

\section{Introduction}

Locality is a cherished principle in physics. Relativistic causality - the fact that no physical information carrying signal can propagate faster than the speed of light - is implemented in the statement that spacelike separated fields commute. This ensures that laboratories in spacelike separated spacetime regions function independently. A deep result expressing this independence in the algebraic formulation of quantum field theory is the split property. As a consequence of the split property, we can specify the state of quantum fields independently on different parts of a Cauchy slice. This lore is being challenged \cite{Laddha:2020kvp,Chowdhury:2020hse,Raju:2020smc,Raju:2021lwh,SuvratYouTube} in the setting of quantum gravity, by the principle of the holography of information, which claims that 

{\vskip 0.25cm}

{\it In a theory of quantum gravity, a copy of all the information available on a Cauchy slice is also available near the boundary of the Cauchy slice. This redundancy in description is already visible in the low-energy theory.}

{\vskip 0.25cm}

\noindent
This principle demands a dramatic revision of intuition built on locality. For example, the principle of holography of information implies that given the state near the boundary of the Cauchy slice, the rest of the state is determined: the split property fails and we are not guaranteed that laboratories in spacelike separated regions of spacetime function independently! The principle of the holography of information is a source of a dramatic new non-locality\footnote{For related studies we refer the interested reader to \cite{tHooft:1993dmi,Susskind:1994vu,Marolf:2008mf,Donnelly:2017jcd,Bousso:2017xyo,Donnelly:2018nbv,Giddings:2020usy,Giddings:2021khn,Bahiru:2022oas}.}. 

The argument of \cite{Laddha:2020kvp,Chowdhury:2020hse,Raju:2020smc,Raju:2021lwh,SuvratYouTube} is compelling in its simplicity. The principle of the holography of information has two basic ingredients. The first is the Reeh-Schlieder Theorem \cite{Reeh:1961ujh}, which is a Theorem about relativistic quantum field theory. We simply state the theorem and refer the reader to \cite{Witten:2018zxz} for a readable account with details. Denote the vacuum of the quantum field theory as $|\Omega\rangle$ and use ${\cal H}_0$ to denote the vacuum sector\footnote{The vacuum sector is not necessarily the full Hilbert space as there may be superselection sectors. This happens, for example, when there are conserved charges that are not carried by any local operator. In a non-trivial superselection sector an analogue of the Reeh-Schlieder theorem holds, so the existence of non-trivial superselection sectors should not distract us.} of the full Hilbert space ${\cal H}$. The vacuum sector consists of all states that can be created from the vacuum by applying local field operators. Assuming\footnote{This assumption is to simplify the discussion and is easily relaxed \cite{Witten:2018zxz}.} that the algebra of local fields is generated by a hermitian scalar field $\phi(x^\mu)$, we introduce a smeared field $\phi_f\equiv\int d\vec{x}f(x^\mu)\phi(x^\mu)$ and a set of states (both $n$ and the functions $f_i$ are varied to get the full set)
\bea
|\Psi_{\{ f_1,\cdots,f_n\}}\rangle =\phi_{f_1}\phi_{f_2}\cdots\phi_{f_n}|\Omega\rangle
\eea
Let $\Sigma$ be a Cauchy hypersurface. Consider an arbitrarily small open set ${\cal V}\subset\Sigma$ and let ${\cal U}_{\cal V}$ be a small neighbourhood of ${\cal V}$ in spacetime. The Reeh-Schleider theorem states that even after restricting the functions $f_i$ to support in ${\cal U}_{\cal V}$, the states $|\Psi_{\{ f_1,\cdots,f_n\}}\rangle$	generate ${\cal H}_0$. This remarkable result reflects the enormous amount of entanglement in the quantum field theory vacuum. The second ingredient that goes into the principle of the holography of information is that, as a consequence of the Gauss law, the energy of a state in gravity can be measured from near the boundary. This implies that the projector onto the state of lowest energy, $P_\Omega =|\Omega\rangle\langle\Omega|$, is an element of the boundary algebra of operators. The principle now follows \cite{Laddha:2020kvp,Chowdhury:2020hse,Raju:2020smc,Raju:2021lwh,SuvratYouTube}: First, note that any observable in ${\cal H}_0$ can be written as a linear combination of operators of the form $|a\rangle\langle b|$ where $|a\rangle$ and $|b\rangle$ are allowed to be any states in ${\cal H}_0$. Using the Reeh-Schleider theorem we know the complete set of these operators can be written in the form
\bea
|a\rangle\langle b|=
\phi_{f^{(a)}_1}\phi_{f_2^{(a)}}\cdots\phi_{f^{(a)}_{n^{(a)}}}|\Omega\rangle
\langle\Omega|\phi_{f^{(b)}_1}\phi_{f_2^{(b)}}\cdots\phi_{f^{(b)}_{n^{(b)}}}
\eea
Since the Gauss law implies that $P_\Omega$ is an element of the boundary algebra of operators, and since a product of operators in the boundary algebra is again an element of the boundary algebra we conclude that the complete set of operators $|a\rangle\langle b|$ belong to the boundary algebra. Consequently, any observable\footnote{This would include operators that one naively thought were localized deep in the bulk of spacetime.} in ${\cal H}_0$ is an element of the boundary algebra of operators and the principle is proved. 

This unusual localization of quantum information in quantum gravity is the focus of this paper. More concretely, the AdS/CFT correspondence gives a non-perturbative definition of quantum gravity on negatively curved spacetimes in the form of a conformal field theory. Our goal in this article is to use the AdS/CFT correspondence to search for signatures of the principle of the holography of information directly in the conformal field theory. Concretely, we use the bilocal holography of the free O$(N)$ model to study these questions in higher spin gravity. The construction and key results of bilocal holography are reviewed in Section \ref{bilocalholography}. The conformal field theory is described using a bilocal collective field. A key formula from Section \ref{bilocalholography} is the mapping (\ref{map1}) which locates the bulk operator, corresponding to a given bilocal operator, in the bulk AdS$_4$ spacetime. We also review the important fact that there is some freedom in the reconstruction of the bulk fields in the conformal field theory. Results establishing the convergence of the operator product expansion in unitary conformal field theories, in Minkowski spacetime, are reviewed in Section \ref{OPEConv}. In Section \ref{OPEHI} we present our central result: the principle of the holography of information, in bilocal holography, can be verified using the operator product expansion. We speculate on how the principle is realized in AdS/CFT, in more general situations, in Section \ref{conclusions}.

A potential point of confusion can be clarified immediately: the reader might wonder if, in the setting of the AdS/CFT correspondence, the holography of information is trivially true. After all, doesn't the statement of the AdS/CFT correspondence, that the dynamics of the bulk is coded into the dynamics of a conformal field theory living on the boundary, imply the holography of information? This is a misunderstanding of the principle. The principle of the holography of information is a statement about the quantum gravity theory itself. The proof of the principle \cite{Laddha:2020kvp,Chowdhury:2020hse,Raju:2020smc,Raju:2021lwh,SuvratYouTube}, as reviewed above, does not invoke AdS/CFT in any way at all, and consequently it also holds (for example) for a theory of quantum gravity in flat spacetime where a holographic dual is not even established. Our goal is to use AdS/CFT to map the principle of the holography of information into a statement about the conformal field theory. This statement should be proved using only conformal field theory methods i.e. without appealing to AdS/CFT or to the holographic gravity dual.  If this succeeds, it provides non-trivial support for the principle. 

Finally, the setting of our study is higher spin gravity which differs in some important ways from usual Einsteinian gravity. The spectrum of higher spin gravity includes not just a massless spin two graviton, but rather there are massless gauge fields for every even integer spin. It is clear that higher spin gravity will not share all the features of Einsteinian gravity and there may be important differences between the two. Nonetheless, we believe that this is a reasonable arena in which to test the holography of information. Higher spin gravity is a quantum theory - so the Reeh-Schlieder theorem applies, and it does enjoy the gauge invariance that is responsible for the Gauss law. Thus the key ingredients needed to prove the principle are present.

\section{Bilocal Holography}\label{bilocalholography}

The AdS/CFT correspondence \cite{Maldacena:1997re,Gubser:1998bc,Witten:1998qj} relates a conformal field theory (with loop expansion parameter $\hbar$) to a theory of quantum gravity (with loop expansion parameter ${1\over N}$). Changing the loop expansion parameter requires a non-trivial rearrangement of the conformal field theory degrees of freedom. It can be achieved by collective field theory \cite{Jevicki:1979mb,Jevicki:1980zg} which expresses the theory in terms of invariant variables. The key insight is that the collective field variables have no explicit $N$ dependence, so that the ${1\over N}$ expansion is manifestly generated as the loop expansion of the collective field theory. Since collective field theory provides a constructive approach to holography, it is the ideal framework for this study. In what follows we work at the leading order in the large $N$ expansion.

The conformal field theory we study, the free O$(N)$ model, has the Lagrangian
\bea
{\cal L}={1\over 2}\partial_\mu\phi^a\partial^\mu\phi^a
\eea
and is defined in $2+1$ dimensions. There is compelling evidence \cite{Giombi:2009wh} that this theory is AdS/CFT dual \cite{Klebanov:2002ja,Sezgin:2002rt} to higher spin gravity \cite{Vasiliev:1990en,Vasiliev:2003ev,Didenko:2014dwa} in AdS$_4$ spacetime. Holography for vector models, using a collective field description, was first proposed in \cite{Das:2003vw} and then developed in a series of papers\footnote{Related but distinct ideas were recently put forward in \cite{Aharony:2020omh}.} \cite{deMelloKoch:2010wdf,Jevicki:2011ss,Jevicki:2014mfa,deMelloKoch:2014mos,deMelloKoch:2014vnt,Jevicki:2015sla,deMelloKoch:2018ivk}, to which the reader is referred for more details. The discussion is most transparently carried out using a lightfront quantization, since it is then possible to choose light cone gauge and to reduce to physical degrees of freedom. Denote the conformal field theory coordinates with little letters as $x^+,x^-,x$ and the coordinates of the dual AdS$_4$ spacetime with capital letters as $X^+,X^-,X,Z$, with $Z$ the extra holographic coordinate. For the O$(N)$ model, at each time $x^+$ we change from the original field $\phi^a(x^+,x^-,x)$ to a new set of gauge invariant variables, given by the bilocal fields
\bea
\sigma (x^+,x_1^-,x_1,x_2^-,x_2)
=\phi^a (x^+,x_1^-,x_1)\phi^a (x^+,x_2^-,x_2)\label{bilocal}
\eea
where the index $a$ is summed. The bilocal packages the complete set of independent single trace equal $x^+$ gauge invariant fields. This collective field is a function of 5 coordinates. In what follows, it is convenient to perform a Fourier transform in the $x^-$ coordinate, which trades $x_1^-$ and $x_2^-$ for the conjugate momenta $p_1^+$ and $p_2^+$. We also perform a Fourier transform in the AdS spacetime, trading coordinate $X^-$ for coordinate $P^+$. The change of field variable from $\phi^a$ to $\sigma$ is associated with a Jacobian which is highly non-linear and leads to an infinite sequence of interaction vertices \cite{deMelloKoch:1996mj}. The single trace spectrum of primary operators includes a scalar of dimension $\Delta=1$ and higher spin currents $J^{\mu_1\cdots\mu_{2s}}$ of every even integer spin $2s$ and dimension $\Delta=2s+1$. As usual, every single trace primary corresponds to a field of the dual higher spin gravity: there is a massless gauge field $A^{M_1\cdots M_{2s}}$ of every even integer spin, as well as a scalar field. The bilocal develops a large $N$ expectation value, which we denote as $\sigma_0(x^+,x_1^-,x_1,x_2^-,x_2)$. Expanding about this background defines the fluctuation $\eta(x^+,x_1^-,x_1,x_2^-,x_2)$ 
\bea
\sigma (x^+,x_1^-,x_1,x_2^-,x_2)=\sigma_0(x^+,x_1^-,x_1,x_2^-,x_2)
+\eta(x^+,x_1^-,x_1,x_2^-,x_2)
\eea
It is the fluctuation $\eta(x^+,x_1^-,x_1,x_2^-,x_2)$ that is identified with the fields of the higher spin gravity. Note that we can write $\eta(x^+,x_1^-,x_1,x_2^-,x_2)=:\phi^a(x^+,x_1^-,x_1)\phi^a(x^+,x_2^-,x_2):$. We will use this equation below.

The higher spin currents of the conformal field theory are traceless and conserved. Consequently, not all components of the current are independent. In the end, there are two independent components of the current at each spin. In the higher spin gravity, we fix light cone gauge $A^{+ M_2\cdots M_{2s}}=0$ and solve the associated constraint. This leave a gauge field with all polarizations transverse to the lightcone $A^{XXZZX\cdots}$. This gauge field is totally symmetric and traceless, so that in the end we have two independent physical components of the gauge field at each spin \cite{Metsaev:1999ui}. The basic claim of bilocal holography is that the theory of the physical degrees of freedom of the higher spin gauge field is given by the bilocal collective field theory description of the independent components of the conformal field theory current. This is supported by the explicit form of the GKPW dictionary, worked out in the lightcone gauge, in \cite{Mintun:2014gua}, and the fact that by performing a suitable change of coordinates the generators of the conformal group \cite{Metsaev:1999ui} acting on the independent components of the currents in the conformal field theory and on the physical degrees of freedom in the gravity, are mapped into each other. To write this representation in the higher spin theory it is useful to employ the four AdS spacetime coordinates as well as an additional variable $\theta$ whose role is to organize the higher spin fields. See equation (\ref{thetamodeexpn}) below. The change of coordinates that relates the conformal field theory and gravity representations  identifies $x^+=X^+$, and relates the remaining conformal field theory coordinates $(p_1^+,x_1,p_2^+,x_2)$ to the remaining AdS coordinates $(P^+,X,Z,\theta)$ as follows
\bea
x_1&=& X+Z \tan \left(\frac{\theta }{2}\right)\qquad
x_2\,\,=\,\, X-Z \cot \left(\frac{\theta }{2}\right)\cr
p_1^+&=& P^+ \cos ^2\left(\frac{\theta }{2}\right)\qquad\quad
p_2^+\,\,=\,\, P^+ \sin ^2\left(\frac{\theta }{2}\right)\label{map1}
\eea
This is easily inverted
\bea
X&=& \frac{p_1^+ x_1+p_2^+ x_2}{p_1^++p_2^+}\qquad
Z\,\,=\,\,\frac{\sqrt{p_1^+ p_2^+} |x_1-x_2|}{p_1^++p_2^+}\cr
P^+&=& p_1^++p_2^+\qquad\qquad
\theta\,\,=\,\,2 \tan ^{-1}\left(\sqrt{\frac{p_2^+}{p_1^+}}\right)\label{map2}
\eea
In addition to mapping the generators correctly, one also finds that the conformal field theory equations of motion for the independent components of the currents are mapped into the higher spin equations of motion for the physical degrees of freedom of the higher spin gauge field. In this way the equivalence between the bilocal collective field theory and the AdS higher spin gravity is made manifest. The utility of this map is that it nicely explains where localized excitations in the conformal field theory map into the bulk. 

One natural application of the above result is to the problem of subregion duality. By studying the subregion $-L\le x \le L$ the paper \cite{deMelloKoch:2021cni} showed that it is possible to reconstruct the bulk region given by $X^2+Z^2\le L^2$. The curve $X^2+Z^2=L^2$ is the geodesic in the AdS$_4$ spacetime that connects the endpoints $X=\pm L$, $Z=0$ on this constant $X^+$ slice, so it is the Ryu-Takayanagi surface \cite{Ryu:2006bv} in lightcone quantization. In this way we recover the expected entanglement wedge reconstruction result, giving confidence that the bilocal collective map encodes the conformal field theory degrees of freedom into the AdS$_4$ bulk in the correct way.

We are interested in the location of operators, after mapping to the dual gravity, in the holographic $Z$ direction. We know that both $p_1^+>0$ and $p_2^+>0$. Consequently in the formula
\bea
Z\,\,=\,\,\frac{\sqrt{p_1^+ p_2^+}}{p_1^++p_2^+}|x_1-x_2|
\eea
the prefactor ${\sqrt{p_1^+ p_2^+}\over p_1^++p_2^+}$ is non-zero and always less than one. By restricting to bilocals with $|x_1-x_2|\le\epsilon$ we restrict to operators that live in a tiny band in the neighbourhood of the boundary, certainly within $Z<\epsilon$. An arbitrarily small $\epsilon$ implies that this band becomes arbitrarily small. Conversely, the only way in which we can probe deep into the bulk, corresponding to large values for $Z$, is by making the separation $x_1-x_2$ large.

It is interesting to ask where single trace primaries map to in the AdS$_4$ bulk. The scalar primary is given by $\phi^a(x^+,x^-,x)\phi^a(x^+,x^-,x)$ i.e. it is obtained from the bilocal (\ref{bilocal}) by setting $x_1=x_2=x$ and $x_1^-=x_2^-=x^-$. Consequently, it is located on the boundary $Z=0$. The conserved currents are given by
\begin{eqnarray}
J_s(x^+,x^-,x,\alpha)
&=&J_{\mu_1\mu_2\cdots\mu_s}(x^+,x^-,x)\alpha^{\mu_1}\alpha^{\mu_2}\cdots \alpha^{\mu_s}\cr\cr
&=&\sum_{k=0}^{s}
\frac{(-1)^k\, :(\alpha\cdot\partial)^{s-k}\phi^a(x^+,x^-,x)\;(\alpha\cdot\partial)^{k}\phi^a(x^+,x^-,x) :}
{k!(s-k)!\Gamma(k+{1\over 2})\Gamma(s-k+{1\over 2})}
\cr
&&\label{scurrent}
\end{eqnarray}
where $\alpha^\mu$ is a polarization vector employed as a convenient book keeping device. The equal $x^+$ bilocal field eliminates components of the current with $+$ polarizations, so we will set $\alpha^+=0$. In this case the derivatives above are all with respect to $x^-$ or $x$. To express these currents in terms of the bilocal field we need to separate the points slightly so that we can act with derivatives on either field separately
\begin{eqnarray}
J_s(x^\mu,\alpha)
&=&\sum_{k=0}^{s}
\frac{(-1)^k\, :(\alpha\cdot\partial_1)^{s-k}\;(\alpha\cdot\partial_2)^{k} :}
{k!(s-k)!\Gamma(k+{1\over 2})\Gamma(s-k+{1\over 2})}
\eta(x^+,x_1^-,x_1,x_2^-,x_2)\Big|_{x_1=x_2=x,x_1^-=x_2^-=x^-}
\cr
&&
\end{eqnarray}
For the purpose of constructing the spinning current it is enough to separate the two points $x_1$ and $x_2$ by an arbitrarily small amount $\epsilon$, evaluate the relevant derivatives and then send $x_2\to x_1$. Put differently, we can construct the current at $x_1$ from the bilocal field $\eta(x^+,x_1^-,x_1,x_2^-,x_2)$ with $|x_1-x_2|<\epsilon$ where $\epsilon$ can be arbitrarily small. Thus the complete set of single trace primary operators, after mapping to the dual gravity, are supported in an arbitrarily small neighbourhood of the boundary.

\begin{figure}[h]%
\begin{center}
\includegraphics[width=0.48\columnwidth]{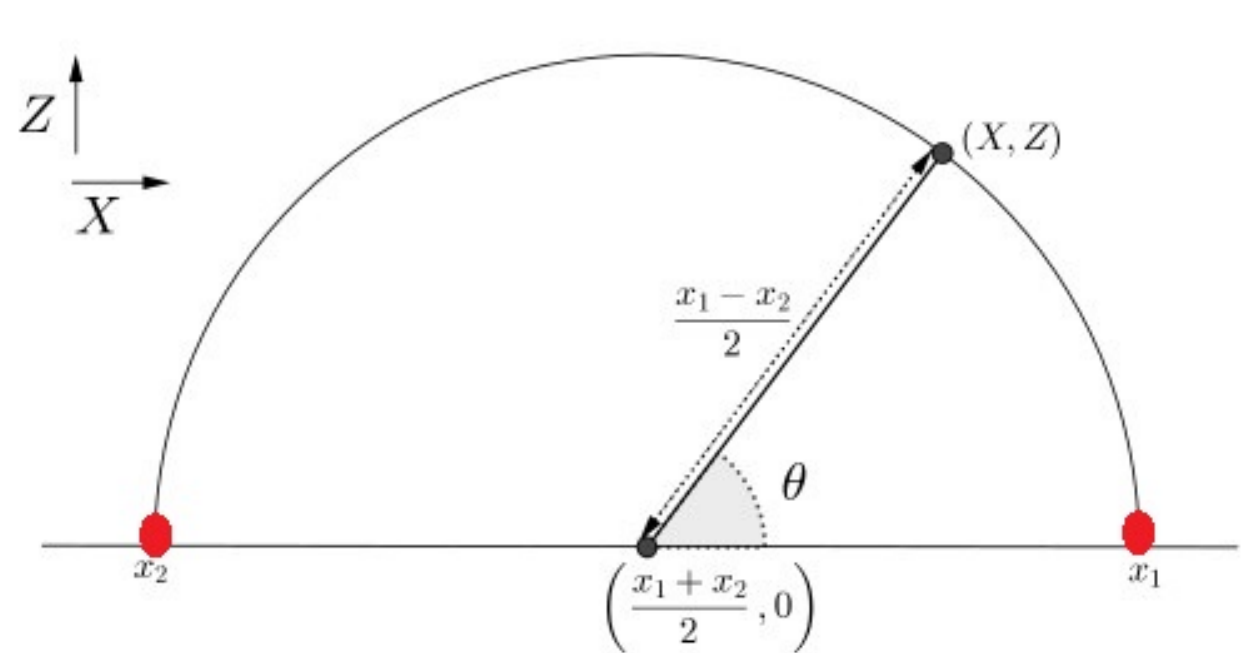}%
\caption{The bilocal describing a pair of excitations localized at $(x_1,p_1^+)$ and $(x_2,p_2^+)$ correspond to a 
bulk excitation localized at $(X,Z)$ as shown.
The bilocal conformal field theory excitation (two red circles) maps into a bulk excitation on the semicircle above.
This figure lives on a constant $x^+=X^+$ slice.
The angle $\theta$ is related to $p_1^+$ and $p_2^+$ according to (\ref{ForTheta}).}
\label{fig:ToBulk}
\end{center}
\end{figure}

\begin{figure}[h]%
\begin{center}
\includegraphics[width=0.65\columnwidth]{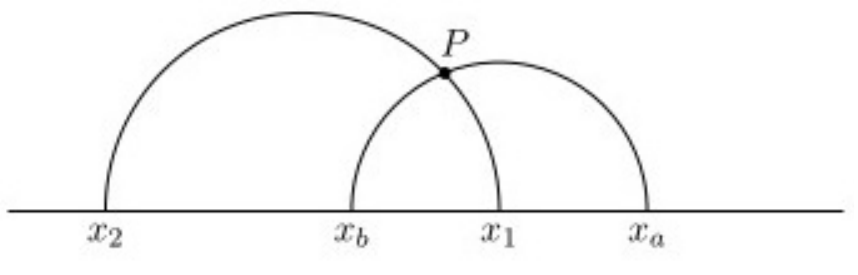}%
\caption{The bulk field located at bulk point $P$ can be constructed using the bilocal $\sigma(x^+,p_1^+,x_1,p_2^+,x_2)$ or using the bilocal $\sigma(x^+,p_1^+,x_a,p_2^+,x_b)$ due to fluidity in the reconstruction of bulk fields from bilocals in the conformal field theory. There are an infinite number of different possible reconstructions corresponding to the fact that an infinite number of semi-circles with distinct endpoints, all passing through $P$, can be drawn.}
\label{fig:QEC}
\end{center}
\end{figure}

Finally, it was pointed out in \cite{deMelloKoch:2021cni} there is some freedom in the reconstruction of the bulk fields. This freedom will be used in what follows. To see how this arises, notice that from the map (\ref{map1}) and (\ref{map2}) it follows that a bilocal with coordinates $x_1$ and $x_2$ maps to a semi-circle
\bea
\left( X-{x_1+x_2\over 2}\right)^2+Z^2=\left({x_1-x_2\over 2}\right)^2
\eea
in the bulk. Some simple trigonometry (see Figure \ref{fig:ToBulk}) implies that
\bea
\tan\theta={Z\over X-{x_1+x_2\over 2}}={2\sqrt{p_1^+p_2^+}\over p_1^++p_2^+}
\label{ForTheta}
\eea

\noindent
so that $\theta$ appearing in Figure \ref{fig:ToBulk} is the same coordinate $\theta$ appearing in the map \cite{deMelloKoch:2021cni}. Consequently, varying $p_1^+$ and $p_2^+$ moves us along the semi-circle. To obtain a definite value of $\theta$ and $P^+$ for the bulk field, we need to evaluate the bilocal at
\bea
p_1^+= {1\over 2}P^+\left(1-\cos\theta\right)\qquad
p_2^+={1\over 2}P^+\left(1+\cos\theta\right)
\eea
We can then Fourier transform to obtain a field localized at a definite $X^-$, if we wish to. The bulk field reconstructed in this way is a linear combination of many different spinning gauge fields
\bea
\Phi(X^+,X^-,X,Z,\theta)=\sum_{s=-\infty}^\infty 
\left( A^{ZZ\cdots ZZ}\cos (2s\theta)+A^{ZZ\cdots ZX}\sin (2s\theta)\right)
\eea
Components of the gauge field with additional $x$ polarizations immediately follow from the fact that the gauge field is completely symmetric and traceless. To obtain a specific gauge field of a definite spin, localized at a specific bulk point, we need to do an integral over $\theta$ as follows
\bea
A^{ZZ\cdots ZZ}&=&\int_0^\pi d\theta\,\, \Phi(X^+,X^-,X,Z,\theta)\cos (2s\theta)
\cr\cr\cr
A^{ZZ\cdots ZX}&=&\int_0^\pi d\theta\,\, \Phi(X^+,X^-,X,Z,\theta)\sin (2s\theta)
\label{thetamodeexpn}
\eea
Thus, a gauge field with a definite spin, a definite polarization and located at a definite bulk point in AdS$_4$ comes from a bilocal located at a definite $x_1,x_2,{x_1^-+x_2^-\over 2}$ but completely smeared over the relative coordinate $x_1^--x_2^-$. Further, for this construction we can use any semi-circle that passes through the bulk point so that infinitely many different reconstructions of the bulk field, each using a different bilocal field\footnote{The different bilocals have different values of $x_1$ and $x_2$. See Figure \ref{fig:QEC}.}, are possible. This is the fluidity in the bulk/boundary dictionary that appears in the quantum error correction interpretation of AdS/CFT \cite{Almheiri:2014lwa}.  This fluidity is needed to resolve apparent inconsistencies of bulk reconstruction.

Although we do not need it for what follows, note that bilocal holography has been studied for the interacting IR fixed point of the $O(N)$ model in \cite{Mulokwe:2018czu,Johnson:2022cbe} and the bilocal holography of the thermofield double has been constructed in \cite{Jevicki:2015sla,Jevicki:2015pza,Jevicki:2021ddf}.

\section{Convergence of the Operator Product Expansion}\label{OPEConv}

In the next section we will be making use of the operator product expansion (OPE). To prepare for this application, this section reviews results about the convergence of the OPE in unitary conformal field theories, in Minkowski spacetime.

In conformal field theory the OPE, which expresses the product of two fields at different points as a sum of a (possibly infinite) number of local fields, is often a convergent expansion. Our application uses the OPE of two identical scalar operators. Conformal symmetry groups all local operators of the theory into conformal multiplets, consisting of a primary operator together with its derivatives (descendants). The OPE is written in terms of a sum over the primary operators ${\cal O}$ as follows
\bea
\sum_{a=1}^N\phi^a(x^\mu+y^\mu)\phi^a(x^\mu-y^\mu) =\sum_{\cal O}
f_{\phi\phi {\cal O}}P_{\cal O}(y^\mu, \partial_x){\cal O}(x)\label{OPEagain}
\eea
If the primary operators ${\cal O}$ have a non-zero spin they will also have indices. The contractions of these indices is not written explicitly above. The coefficient function $P_{\cal O}$ is a power series in $\partial_y$ which encodes the contribution of the primary ${\cal O}$ and all of its descendants. The form of this function is completely fixed by conformal invariance in terms of the operator scaling dimensions. The number $f_{\phi\phi{\cal O}}$ is called the OPE coefficient and it together with the spectrum of scaling dimensions of the primary operators completely determines the dynamical content of the conformal field theory. 

The convergence of the OPE is established by using it to relate an $n+2$ point correlation function to an $n+1$ point correlation function as follows \cite{Pappadopulo:2012jk}
\bea
\langle\phi^a (x)\phi^a (y)\prod_{i=1}^n\psi_i (z_i)\rangle
=\sum_{{\cal O}} f_{\phi\phi{\cal O}}P_{\cal O}(x-y,\partial_y)
\langle {\cal O}(y)\prod_{i=1}^n\psi_i (z_i)\rangle\label{CrFn}
\eea
There is some freedom in writing this formula as we might write the product $\phi (x)\phi (y)$ as a sum of operators located at different points. Natural choices include at $x$, at $y$ or at the midpoint $(x+y)/2$. The statement that the OPE converges is that statement that the right hand side of the above formula is absolutely convergent at finite separation $x-y$, rather than being just an asymptotic expansion in the limit $x\to y$. It is simplest to start in Euclidean space where we can appeal to radial quantization. The convergence of the OPE expansion is then related to the convergence of a scalar product of two Hilbert space states. The argument \cite{Polchinski:1998rq} starts by quantizing the theory radially with point $y$ as the origin. In the case that
\bea
|x-y| < \min_i |z_i-y|
\eea
we can find a sphere separating the points $x,y$ from the points $z_i$ where the remaining operators are inserted. The LHS of (\ref{CrFn}) is then the overlap $\langle\Psi|\Phi\rangle$ of the two states
\bea
|\Phi\rangle =\phi^a(x)\phi^a(y)|0\rangle\qquad 
\langle\Psi|=\langle 0|\prod_{i=1}^n\psi_i (z_i)
\eea
produced by acting on the radial quantization in and out vacua. Thus, the convergence of the OPE expansion is related to the convergence of a scalar product of two Hilbert space states in radial quantization. Convergence is then implied by a basic theorem about Hilbert spaces: the scalar product of two states converges when one of the two states is expanded into an orthonormal basis.

What is the rate of convergence of the OPE? By focusing on four point functions, this rate was studied in \cite{Pappadopulo:2012jk}. The expansions are convergent in a finite region with an exponential convergence rate for the two different schemes considered, corresponding to the case that $\phi(x)\phi(y)$ is expressed as a sum of operators inserted at $y$ or at ${x+y \over 2}$ \cite{Pappadopulo:2012jk}. All results described so far refer to the convergence of the OPE in the Euclidean theory. To obtain convergence results for the Minkowskian theory, the Euclidean four point functions need to be analytically continued to imaginary time. This question has been considered carefully in the paper \cite{Qiao:2020bcs} which studied the convergence properties of operator product expansions (OPE) for Lorentzian conformal field theory four-point functions of scalar operators implied by analytic continuation of the Euclidean results we have just reviewed. The key results for us are Theorem 4.1, Theorem 4.4 and Theorem 4.6 of \cite{Qiao:2020bcs} which give the criteria for the convergence of the $s$-channel, $t$-channel and $u$-channel OPEs. In terms of the conformal cross ratios
\bea
u={x_{12}^2 x_{34}^2\over x_{13}^2 x_{24}^2}\qquad
v={x_{14}^2 x_{23}^2\over x_{13}^2 x_{24}^2}
\eea
where $x^\mu=(t,\vec{x})$ and
\bea
x_{ij}^2=-(t_i-t_j)^2+(\vec{x}_i-\vec{x}_j)\cdot (\vec{x}_i-\vec{x}_j)
\eea
we introduce two parameters $z,\bar{z}$ as follows
\bea
u=z\bar{z}\qquad v=(1-z)(1-\bar{z})
\eea
If neither $z$ nor $\bar{z}$ belong to $(1,\infty)$ then the $s$-channel OPE is convergent, if neither $z$ nor $\bar{z}$ belong to $(-\infty,0)$ then the $t$-channel OPE is convergent and finally if neither $z$ nor $\bar{z}$ belong to $(0,1)$ then the $u$-channel OPE is convergent. These are the basic results that we use below. 

We will be using the OPE for products of operators that all live on the same equal $x^+$ slice. In this case
\bea
x_i^\mu = (x^+,x^-_i,x_i)\qquad x_{ij}^2=(x_i-x_j)^2
\eea
so that only the coordinate transverse to the lightcone appears in the conformal cross ratios. The parameters $z,\bar{z}$ are then defined by
\bea
{(x_1-x_2)^2 (x_3-x_4)^2\over (x_1-x_3)^2(x_2-x_4)^2}=z\bar{z}\qquad 
{(x_1-x_4)^2 (x_3-x_2)^2\over (x_1-x_3)^2(x_2-x_4)^2}=(1-z)(1-\bar{z})
\eea

\begin{figure}[h]%
\begin{center}
\includegraphics[width=0.65\columnwidth]{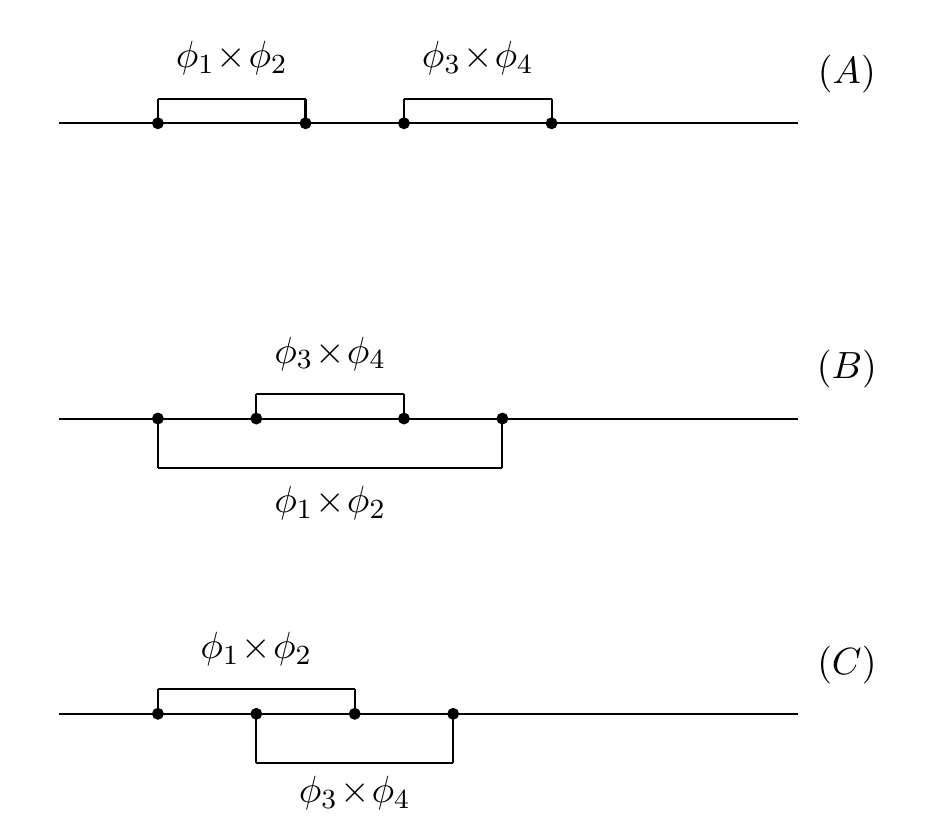}
\caption{The $s$-channel OPE converges for the configurations (A) and (B) above. It does not converge for configuration (C).}
\label{fig:OPEconvConfg}
\end{center}
\end{figure}
In terms of these parameters, the conditions for the convergence of the OPE are rather intuitive, see Figure \ref{fig:OPEconvConfg}. For the purpose of illustration, consider the $s$-channel OPE. The s-channel OPE computes the operator products $\phi_1\times\phi_2$ and $\phi_3\times\phi_4$. For example, consider the four point function
\bea
&&\langle\sigma (x^+,x_1^-,x_1,x_2^-,x_2)
\sigma (x^+,x_3^-,x_3,x_4^-,x_4)\rangle\cr
&&\qquad
=\langle \phi^a(x^+,x_1^-,x_1)\phi^a (x^+,x_2^-,x_2)
\phi^b (x^+,x_3^-,x_3)\phi^b(x^+,x_4^-,x_4)\rangle
\eea
and choose $(x_1,x_2,x_3,x_4)=(1,2,3,4)$ which gives a value of
\bea
 z={1\over 4}=\bar{z}
\eea
According to \cite{Qiao:2020bcs} the $s$-channel OPE expressing each bilocal as a sum of local operators converges. This is also the case for the choice $(x_1,x_2,x_3,x_4)=(5,2,3,4)$ which corresponds to 
\bea
 z={3\over 4}=\bar{z}
\eea
If on the other hand we choose $(x_1,x_2,x_3,x_4)=(1,3,2,4)$ we find a situation in which the two bilocals ``straddle'' each other and we do not expect the $s$-channel OPE for each bilocal to converge. In this case we find that
\bea
\qquad z=4=\bar{z}
\eea
so that according to \cite{Qiao:2020bcs} the $s$-channel OPE expressing each bilocal as a sum of local operators does not converge. In this case however, we can still use the $t$-channel OPE which computes $\phi_1\times\phi_4$ and $\phi_2\times\phi_3$. This corresponds to a rather non-trivial rearrangement of the gauge invariant degrees of freedom as this OPE channel takes the product of fields belonging to different gauge invariant bilocal fields. There is however nothing wrong with proceeding in this way and in fact equality of these channels has been used for the conformal bootstrap in \cite{Kos:2013tga}.

Finally, note that the only difference between the OPE $\phi^a(x_1)\phi^a(x_2)$ and the OPE $:\phi^a(x_1)\phi^a(x_2):$ is that we do not include the contribution of the identity operator in $:\phi^a(x_1)\phi^a(x_2):$. The identity operator is a scalar primary that has no descendants. The contribution of the identity corresponds to the contribution coming from contracting $\phi^a(x_1)$ with $\phi^a(x_2)$, which is precisely what the normal ordering subtracts out.

\section{OPE and Holography of Information}\label{OPEHI}

The principle of the holography of information predicts that any collection of bulk operators can be expressed as an element of the boundary algebra. In Section \ref{bilocalholography} we have argued that all of the single trace primary operators are supported in an arbitrarily small neighbourhood of the boundary. By separating $x_1$ and $x_2$ to be arbitrarily distant, the bilocal field $\eta(x^+,x^-_1,x_1,x^-_2,x_2)$ corresponds to a bulk operator located arbitrarily deep in the bulk, i.e. it is located at an arbitrarily large $Z$ value in the AdS$_4$ bulk. Thus, the holography of information is verified if we could replace the bilocal field by a sum of single trace primaries. This is exactly what the OPE does, so we see that in the conformal field theory, the principle of the holography of information has reduced to the statement of the OPE. This proves that measuring an appropriate linear combination of operators that belong to the boundary algebra is equivalent to measuring the bulk operator. 

This argument glosses over an important point: the radius of convergence of the OPE inside a correlator is not predetermined but depends on the next-closest operator insertion. Thus, we might spoil the convergence of the OPE for any given bilocal, by including another bilocal that straddles it, exactly as in Figure \ref{fig:OPEconvConfg} (C). To really turn the above observation into a careful argument, one would need to show that this bilocal overlapping problem can always be avoided, i.e. that it never prevents us from writing any product of bulk operators as a convergent sum of gauge invariant operators belonging to the boundary algebra. For the case of a single bulk field, corresponding to a single bilocal, this is indeed the case. When more than one bulk field acts, more care is needed. We start by considering the action of two bulk fields and then three bulk fields before considering the general case.

\subsection{Two bulk fields acting}

In the case of two bulk fields acting there is, apparently, already the possibility that convergence of the OPE is spoiled. For a product of bilocals
\bea
\langle \eta (x^+,x^-_1,x_1,x_2^-,x_2)\eta (x^+,x^-_3,x_3,x_4^-,x_4)\rangle
\eea
we want to use the OPE in the $s$-channel since both products $:\phi_1\times\phi_2:$ and $:\phi_3\times\phi_4:$ are separately gauge invariant. This is not the case for either the $t$ or $u$-channels, so that working in either of these channels we should not expect the result of the OPE to be expressed in terms of single trace primaries and their descendants. 
\begin{figure}[h]%
\begin{center}
\includegraphics[width=0.55\columnwidth]{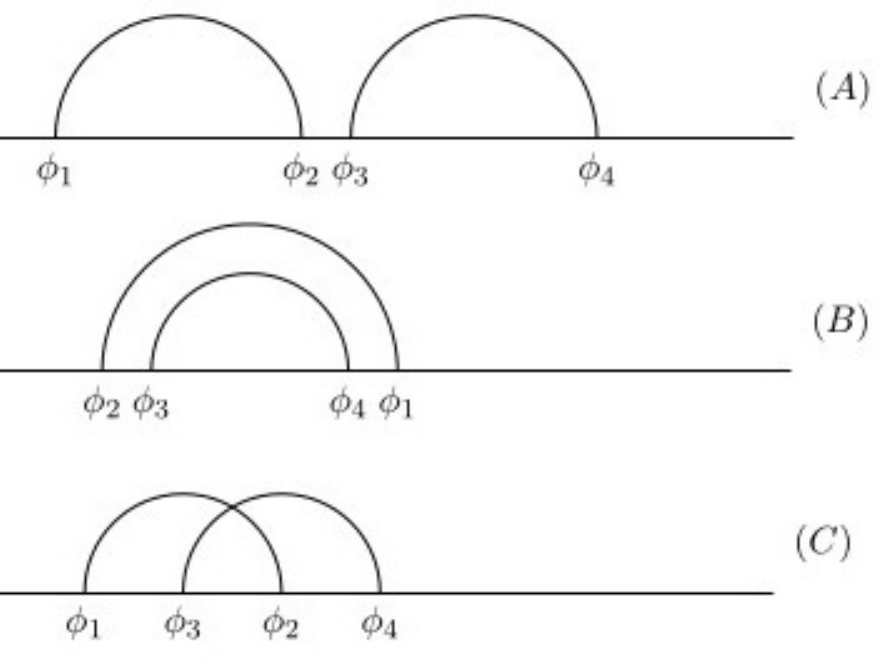}%
\caption{The OPE converges for configurations (A) and (B) and does not converge for configuration (C). In (C) the semicircles depicting the bulk locus of each bilocal intersect.}
\label{fig:OPEsemi}
\end{center}
\end{figure}

However, a potential problem becomes apparent: there are three possible configurations as shown in Figure \ref{fig:OPEsemi}. The $s$-channel OPE converges only for (A) and (B). If configuration (C) arises, we are not able to rewrite each bilocal as a sum of single trace primaries. Fortunately, as we now explain, it is always possible to avoid configuration (C).

The key physical input from bilocal holography that we exploit is the fluidity of the bulk/boundary dictionary. Recall that to reconstruct a field at a bulk point, we can use any bilocal associated to a semi-circle that passes through the point. So the question of whether it is possible or not to avoid configuration (C) boils down to the question of whether or not it is possible to choose semi-circles, each passing through a distinct bulk point, without intersecting each other. To simplify the question, assume that the semi-circles share the same centre i.e. it is only the radius of each semi-circle that changes. The only parameter that we vary when choosing the two semi-circles is their common centre. Since they share a common centre, as long as the radii of the semi-circles are distinct,  they do not overlap. Obviously it is always possible to choose a point on the boundary that is not equidistant from the two bulk points so that, thanks to the fluidity in the bulk reconstruction, we can always arrange to be in configuration (B). Thus, the product of any two bulk operators (acting at distinct events) can always be expressed as an appropriate linear combination of single trace primary operators and their descendants. An observer measuring only observables belonging to the boundary algebra can indeed learn the result that would be obtained by measuring the product of any two bulk operators.

In the Appendix \ref{OPEexamples} we explicitly test the convergence of the OPE in the free $O(N)$ model conformal field theory, for the three configurations shown in Figure \ref{fig:OPEsemi}.

\subsection{Three bulk fields acting}

The case of three bulk fields acting is indicative of the generic bulk configuration. In this case it is no longer possible, for every configuration that can arise, to replace each bilocal with a sum over single trace primaries and their descendants. In general, rewriting the bulk operators in terms of boundary operators necessarily involves scrambling up the information contained in different bilocals. This does not affect the  conclusion that it is possible to write any product of three bulk operators (acting at distinct events) as a convergent sum over operators in the boundary algebra.
 
\begin{figure}[h]
\begin{center}
\includegraphics[width=0.9\columnwidth]{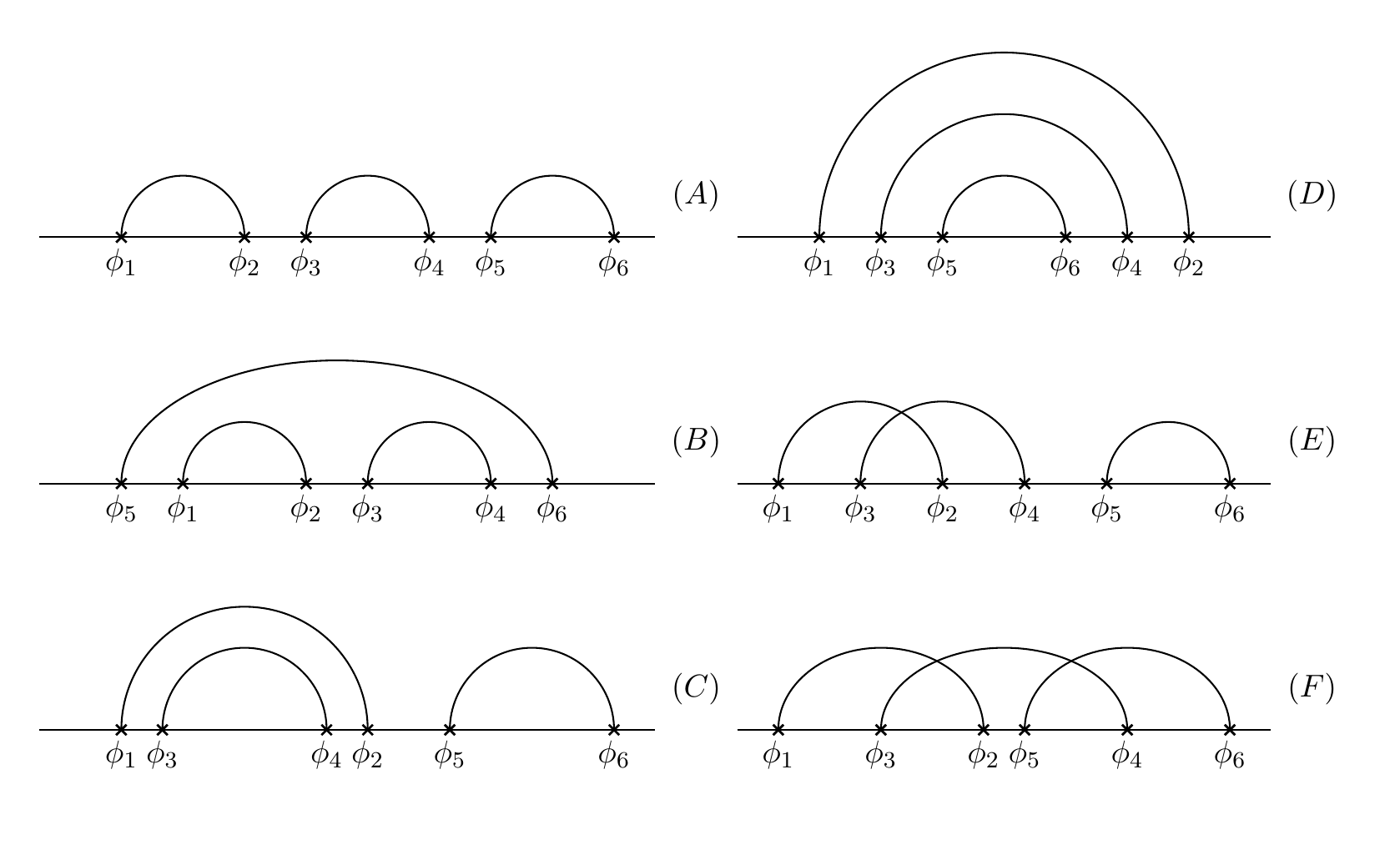}%
\caption{Configurations that arise when three bulk operators act.}
\label{threeact}
\end{center}
\end{figure}

See Figure \ref{threeact} for configurations that are possible when three operators act. For configurations (A) and (B) the OPE channel which computes $\phi_1\times\phi_2$, $\phi_3\times\phi_4$ and $\phi_5\times\phi_6$ converges, so that for these two configrations each bulk operator is individually replaced by a sum over operators localized in the neighbourhood of the boundary. This channel does not converge for the remaining four configurations.

Now, consider configuration (C). An OPE channel that converges computes $\phi_1\times\phi_3$, $\phi_2\times\phi_4$ and $\phi_5\times\phi_6$. It is not obviously possible to express the results of these OPEs in terms of local single trace operators, since both $\phi_1\times\phi_3$ and $\phi_2\times\phi_4$ are not gauge invariant. An extra step is needed, which computes the OPE between the result produced from the $\phi_1\times\phi_3$ OPE and the result produced from the $\phi_2\times\phi_4$ OPE. We denote this using the schematic but transparent notation $(\phi_1\times\phi_3)\times(\phi_2\times\phi_4)$. The result of this final OPE is now gauge invariant and consequently can be expressed in terms of single trace local operators\footnote{We are using the fact that the single trace local operators are a generating set for the complete set of gauge invariant operators. We are implicitly assuming that we consider operators whose dimension is held fixed as $N\to\infty$ to avoid the appearance of new gauge invariant operators constructed using $\epsilon^{a_1\cdots a_N}$. }. Thus, this configuration of bulk operators acting can again be expressed as a sum of local single trace primaries and their products, and thus is an element of the boundary algebra. To achieve this however, we have had to scramble up the information contained in the bilocals $\eta (x^+,x^-_1,x_1,x^-_2,x_2)$ and $\eta (x^+,x^-_3,x_3,x^-_4,x_4)$. The same sequence of OPE computations can be applied to configurations (D) and (E). Finally, for configuration (F) the information mixing between the different bilocals, i.e. between the different bulk fields, is maximal. Indeed, in this case we would need to compute $((\phi_1\times\phi_3)\times (\phi_2\times\phi_5))\times (\phi_4\times\phi_6)$. The arguments of the following section show that we can always choose to avoid this configuration, if we so wish. Indeed, it is always possible to arrange that we have configuration (D).

\subsection{Generic bulk observables}

\begin{figure}[h]%
\begin{center}
\includegraphics[width=0.65\columnwidth]{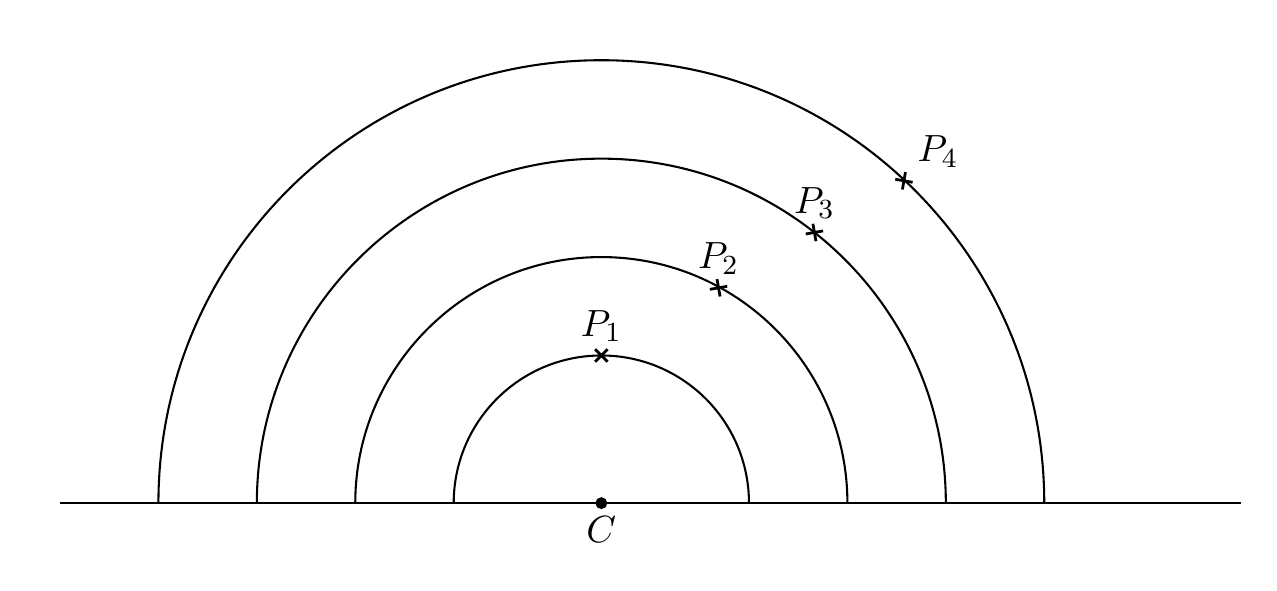}%
\caption{It is always possible to choose a centre ($C$) for $K$ non-intersecting semicircles such that a bulk point $P_i$ lies on the $i$th semicircle. At each semi-circle endpoint there is an operator, which is contracted with the operator at the other endpoint.}
\label{nooverlap}
\end{center}
\end{figure}

Consider the situation in which we have a total of $K$ bulk operators acting at $K$ distinct points in the bulk AdS$_4$ spacetime. We have already seen the each bilocal is associated to a semicircle in the bulk and that any bulk operator located at an event lying on the semicircle can be constructed from the bilocal. We will now argue that it is always possible to choose $p$ semi-circles with the same centre, each passing through a distinct bulk point and each with a distinct radius. For an illustration of a configuration of this type see Figure \ref{nooverlap} for an illustration, with $K=4$. Our configuration of $K$ bulk operators has operators localized at points $(X_i,Z_i,X^-_i)$ $i=1,...,K$. Denote the coordinate of the common centre of the semi-circles as $(X_c,0)$. The condition for points $i$ and $j$ to be equidistant from the centre is
\bea
(X_i-X_c)^2+Z_i^2&=&(X_j-X_c)^2+Z_j^2
\eea
which has a unique solution
\bea
X_c={X_i^2-X_j^2+Z_i^2-Z_j^2\over 2(X_i-X_j)}
\eea
There are thus at most $K(K-1)/2$ values for $X_c$ at which two radii coincide. Choosing any other value for $X_c$ gives a suitable set of semi-circles, which completes the argument.

Moving from left to right the fields are labelled as $\phi_1,\phi_2,\cdots,\phi_{2K-1},\phi_{2K}$. The structure of the sequence of OPE's we compute are dictated by the interplay between gauge invariance and convergence of the OPE, since we need the OPE to converge, and it is only gauge invariant products that are a sum of local single trace operators and their products. We use the product of a pair of boundary observables to describe this collection of bulk operators
\bea
{\cal O}_{\rm boundary}={\cal O}_{\rm boundary,1}{\cal O}_{\rm boundary,2}
\eea
The first boundary observable is obtained by taking a suitable sequence of OPEs so that we can replace $\phi_1,\phi_2,\cdots,\phi_{2K-3},\phi_{2K-2}$ by a local operator
\bea
{\cal O}_{\rm boundary,1}=(\cdots
(((\phi_1\times\phi_2)\times (\phi_3\times\phi_4))\times (\phi_5\times\phi_6))\cdots)
\eea
The second boundary observable computes the OPE $\phi_{2K-1}\times\phi_{2K}$
\bea
{\cal O}_{\rm boundary,2}=(\phi_{2K-1}\times\phi_{2K})
\eea
We have combined $\phi_1,\phi_2,\cdots,\phi_{2K-2}$ into a gauge invariant local operator, through the use of multiple OPEs, and we have combined $\phi_{2K-1},\phi_{2K}$ into a local gauge invariant operator. These are separately both elements of the boundary algebra.

\begin{figure}[h]%
\begin{center}
\includegraphics[width=0.9\columnwidth]{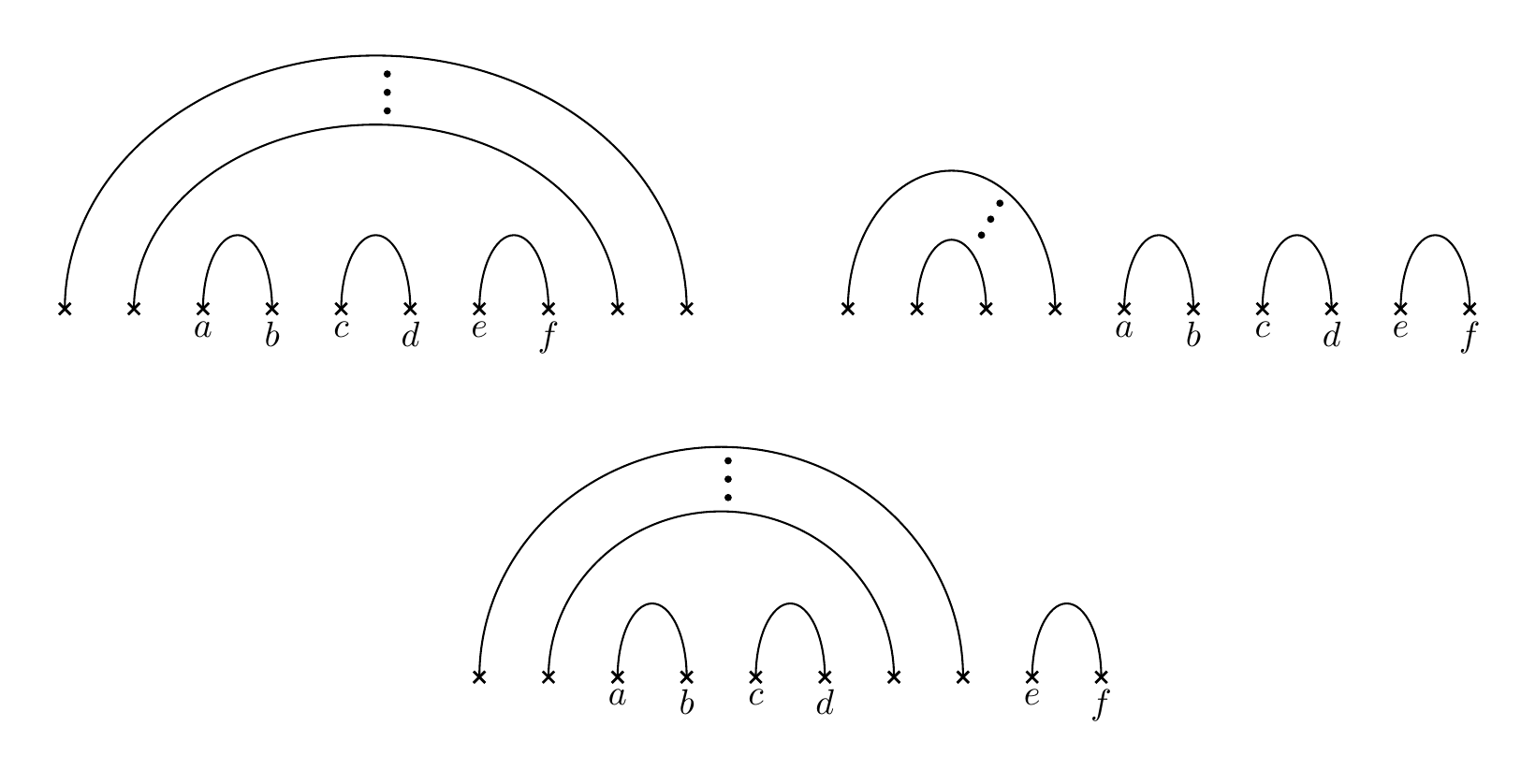}%
\caption{The three smaller semicircles above (with endpoints labelled $a,b,c,d,e,f$) represent small excitations of the bulk state. The bulk state is produced by the bilocals corresponding to the large semicircles. For the first two configurations we can excite the operator ${\cal O}_{\rm boundary}$ itself i.e. the equality of the product of bulk operators and ${\cal O}_{\rm boundary}$ can be used as an operator equation. For the last configuration shown we need to obtain a new representation before we can replace the product of bulk operators by operators living in the boundary algebra.}
\label{lastno}
\end{center}
\end{figure}

To get some insight into the meaning of ${\cal O}_{\rm boundary}$ note that if we excite the bulk state from which ${\cal O}_{\rm boundary}$ was derived, with bilocals that have both points inside all semicircles as shown in the first configuration in Figure \ref{lastno} then we have
\bea
&&\langle \prod_i \eta(x^+,x_i^-,x_i,x_{2K-i}^-,x_{2K-i})
\eta(x^+,x_a^-,x_a,x_b^-,x_b)\eta(x^+,x_c^-,x_c,x_d^-,x_d)\eta(x^+,x_e^-,x_e,x_f^-,x_f)\rangle\cr\cr
&&\qquad=\langle {\cal O}_{\rm boundary,1} \eta(x^+,x_{2K-1}^-,x_{2K-1},x_{2K}^-,x_{2K})\cr\cr
&&\qquad\quad\times\eta(x^+,x_a^-,x_a,x_b^-,x_b)\eta(x^+,x_c^-,x_c,x_d^-,x_d)\eta(x^+,x_e^-,x_e,x_f^-,x_f)\rangle\cr\cr
&&\qquad=\langle {\cal O}_{\rm boundary,1}\tilde{\cal O}_{\rm boundary,2}\rangle
\eea
with $\tilde{\cal O}_{\rm boundary,2}$ a new local operator, arising from a sequence of OPEs between the fields in the bilocal $\eta(x^+,x_{2K-1}^-,x_{2K-1},x_{2K}^-,x_{2K})$ and the fields in the excitations. This gives some insight into the relation between the product of bulk operators and ${\cal O}_{\rm boundary}$: the relation between ${\cal O}_{\rm boundary}$ and the collection of bulk operators can not be treated as an operator equation. As a general conclusion, new excitations are added to a given bulk state a new representation for  ${\cal O}_{\rm boundary}$ must be worked out. This new representation scrambles up the information residing in the excitations and the information in the bulk operators which produced the bulk state. Nevertheless, the conclusion that the collection of bulk operators including the additional excitations, can be represented by an operator belonging to the boundary algebra holds.

\section{Discussion and Conclusions}\label{conclusions}

We have been able to demonstrate the principle of the holography of information in the setting of bilocal holography. The detailed form of the map (\ref{map2}) and the redundancy it implies in reconstructing bulk operators has played an important role in the argument, so it is worth discussing the logic that leads to (\ref{map2}). First collective field theory is used to set up the field theory of invariant variables. The claim is that the collective field theory of the invariant variables is the dual gravitational theory\footnote{Specifically, we do not introduce a mapping involving an integral over a product of a conformal field theory primary field with a kernel that is a Green's function of a linearised bulk equation of motion, to obtain the bulk field. The bulk field is the collective field. We perform the change of variables (\ref{map2}) to make this interpretation explicit.}. In the case of bilocal holography this assertion is supported \cite{deMelloKoch:2010wdf} by showing that the AdS isometry generators and the conformal field theory generators are obtained from each other through the change of variables (\ref{map1}) and (\ref{map2}). The same change of variables relates the equations of motion of the conformal field theory and the higher spin gravity. The fact that this is even possible is not at all obvious given that there are 10 generators in the so(2,3) algebra, and the action of these on an infinite number of conformal field theory currents is to match the action on an infinite number of higher spin gauge fields. With this map in hand the principle of the holography of information is implied by a familiar but remarkable statement in conformal field theory: the operator product expansion. This convincingly confirms the principle of the holography of information and is simultaneously another positive indication that bilocal holography is indeed constructing the quantum gravity dual to the original conformal field theory.

Our discussion has been in the context of the bilocal description of the O$(N)$ vector model, which is dual to higher spin gravity. However, we suspect that these are general lessons about how the principle of the holography of information arises from the conformal field theory, in AdS/CFT. In the case of the vector model, the invariant variables are given by the gauge invariant contraction of a pair of vector fields and thus the invariant fields are bilocal. For theories with matrix valued fields, there are many more ways in which a gauge invariant variable can be constructed. By taking products of matrix fields, located at different points in spacetime and dressed with the necessary Wilson lines to produce a gauge invariant operator, we are naturally lead to bilocal, trilocal and in general multi-local operators. The scale-radius duality of AdS/CFT \cite{Susskind:1998dq} suggests that multi-local operators with well separated locations explore deep into the holographic direction, while the local limit in which all the points in the multi-local operator approach each other, map into a bulk operator located at the boundary. In this case too, the OPE can be used to take the product of separated operators and express them in terms of local operators, so that once again we start to see how all of the information on a given Cauchy slice might be coded into the boundary of that slice.

\begin{center} 
{\bf Acknowledgements}
\end{center}
This work is supported by the South African Research Chairs initiative of the Department of Science and Technology and the National Research Foundation. We thank Antal Jevicki and Suvrat Raju for helpful discussions and Suvrat Raju for patient, generous and extremely helpful feedback on drafts of this paper. This work was completed while the RdMK was a participant of the KITP program ``Integrability in String, Field, and Condensed Matter Theory.'' This research was supported in part by the National Science Foundation under Grant No. NSF PHY-1748958.

\begin{appendix}

\section{OPE Observations}\label{ope}

In this Appendix we work out the operator product expansion for a gauge invariant product of two scalar fields. Exactly as in \cite{Craigie:1983fb} this amounts to an application of Taylor expansions. We use the formulas below to explicitly illustrate the convergence criteria derived in \cite{Qiao:2020bcs}. The primary operators that appear in this operator product are the conserved higher spin currents of spin $2s$, given by (see for example \cite{Giombi:2016hkj})
\bea 
J_{2s}(y,x)&=&y^{\mu_1}\cdots y^{\mu_{2s}} J_{\mu_1\cdots\mu_{2s}}(x)\cr\cr
&=&\pi \sum_{a=1}^N\sum_{k=0}^{2s}(-1)^k
{:\left(y^\mu {\partial\over\partial x^\mu}\right)^{2s-k}\phi^a (x)
\left(y^\nu {\partial\over\partial x^\nu}\right)^k\phi^a (x):\over
k!(2s-k)!\Gamma (k+{1\over 2})\Gamma(2s-k+{1\over 2})}
\eea
and the spin zero primary $J_0=\phi^a(x)\phi^a(x)$. Consider the unequal time bilocal
\bea
\eta (x_1^\mu,x_2^\mu)=:\phi^a(t_1,\vec{x}_1)\phi^a(t_2,\vec{x}_2):
\eea
Introducing the coordinates 
\bea
x^\mu = {1\over 2} (x^\mu_1+x^\mu_2)\qquad y^\mu = {1\over 2}(x_1^\mu-x_2^\mu)
\eea
so that
\bea
{\partial\over\partial x_1^\mu}={1\over 2}\left({\partial\over\partial x^\mu}
+{\partial\over\partial y^\mu}\right)\qquad
{\partial\over\partial x_2^\mu}={1\over 2}\left({\partial\over\partial x^\mu}
-{\partial\over\partial y^\mu}\right)
\eea
we can expand $\eta (x_1^\mu,x_2^\mu)$ to all orders in $y^\mu$ as follows
\bea
\eta (x_1^\mu,x_2^\mu)
&=&\sum_{a=1}^N:\phi^a(x^\mu+y^\mu)\phi^a(x^\mu-y^\mu):\cr\cr
&=&\sum_{a=1}^N\sum_{r,t=0}^\infty {1\over r! t!}
:\left(y^\mu {\partial\over\partial x^\mu}\right)^{r}\phi^a (x)
\left(-y^\nu {\partial\over\partial x^\nu}\right)^t\phi^a (x):
\eea
The fact that we deal with a real field and that we perform the OPE around the midpoint between the two fields implies that odd powers of derivatives sum to zero leaving only even powers. The operator product expansion of $\phi^a$ with itself includes the currents $J_{2s}$ and $J_0$ as well as their descendants so that
\bea
\sum_{a=1}^N:\phi^a(x^\mu+y^\mu)\phi^a(x^\mu-y^\mu):&=&
\sum_{s=0}^\infty\sum_{d=0}^\infty c_{sd} 
\left(y^\mu {\partial\over\partial x^\mu}\right)^{2d} J_{2s}(y,x)
\label{explicitOPE}
\eea 
The number $c_{sd}$ tells us about the contribution of the level $2d$ descendant of the primary current with spin $2s$. To solve for the $c_{sd}$ we can study the polynomial equation
\bea
\left.\sum_{r,s=0}^\infty {1\over r!}{1\over s!}p_1^r p_2^s\right|_{\rm even}
&=&\left.\sum_{s=0}^\infty\sum_{d=0}^\infty c_{sd}(p_1+p_2)^{2d}
\pi\sum_{k=0}^{2s}{(-1)^kp_1^{2s-k}p_2^k\over k!(2s-k)!\Gamma(k+{1\over 2})\Gamma(2s-k+{1\over 2})} \right|_{\rm even}\nonumber
\eea
Here $p_1^qp_2^k$ stands for $:(y^\mu\partial_{x^\mu})^q\phi^a\,(y^\nu\partial_{x^\nu})^k\phi^a:$ so that from both sides we must keep only terms that are symmetric under swapping $1\leftrightarrow 2$. With the help of mathematica we easily find that
\bea
c_{0d}={1\over 2^{2d} (d!)^2} \qquad {\rm and}\qquad
c_{sd}={(2 s)! (4 s-1)!!\over d! 2^{2 d+4 s-1} (d+2 s)!}\qquad\qquad s>0
\eea
Up to this point we have not made any use of conformal symmetry - we have just performed a Taylor expansion. Using conformal symmetry every local operator of the theory can be classified as either a primary operator, or as a derivative of a primary operator, that is, a descendant. A primary operator and all of its descendants belong to the same irreducible representation so it is natural to rewrite the OPE as a sum over just the primaries, which we will denote by ${\cal O}$
\bea
\sum_{a=1}^N\phi^a(x^\mu+y^\mu)\phi^a(x^\mu-y^\mu) =\sum_{\cal O}
f_{\phi\phi {\cal O}}P_{\cal O}(y^\mu, \partial_x){\cal O}\label{OPEagain}
\eea
If the primary operators ${\cal O}$ have a non-zero spin they will also have indices. The contractions of these indices is not written explicitly. The coefficient function $P_{\cal O}$ is a power series in $\partial_y$ which encodes the contribution of the primary ${\cal O}$ and all of its descendants. The form of this function is completely fixed by conformal invariance in terms of the operator scaling dimensions. The number $f_{\phi\phi{\cal O}}$ is called the OPE coefficient and it together with the spectrum of scaling dimensions of the primary operators completely determines the dynamical content of the conformal field theory. For the $O(N)$ model where we know the complete set of primaries we can write (\ref{OPEagain}) slightly more explicitly as
\bea
\sum_{a=1}^N\phi^a(x^\mu+y^\mu)\phi^a(x^\mu-y^\mu) =\sum_{s=0}^\infty
f_{\phi\phi J_{2s}}P_{J_{2s}}(y^\mu, \partial_x) J_{2s}(y,x)\label{OPEagain}
\eea

In the checks that are performed in the next section, (\ref{explicitOPE}) is perfectly sufficient, and we will not need the more elegant result (\ref{OPEagain}).

Finally, we will also find it useful to make use of the OPE
\bea
\sum_{a=1}^N:\phi^a(x^\mu)\phi^a(x^\mu+y^\mu):&=&
\sum_{s=0}^\infty\sum_{d=0}^\infty \tilde{c}_{sd} 
\left(y^\mu {\partial\over\partial x^\mu}\right)^{d} J_{2s}(y,x)
\label{SecondExplicitOPE}
\eea 
The fact that now both even and odd descendants appear simply reflects the fact that this OPE is less symmetrical than the OPE studied in (\ref{explicitOPE}). Again, with the help of mathematica we find
\bea
\tilde{c}_{0d}={(2d-1)!!\over 2^{d} (d!)^2} \qquad {\rm and}\qquad
\tilde{c}_{sd}={(2 s)! (4 s - 1)!! (2 d + 4 s - 1)!!\over 2^{d + 4 s - 1} (d + 4 s)! d!}\qquad s>0
\eea

\subsection{Instructive examples of OPE convergence}\label{OPEexamples}

\begin{figure}[h]%
\begin{center}
\includegraphics[width=0.65\columnwidth]{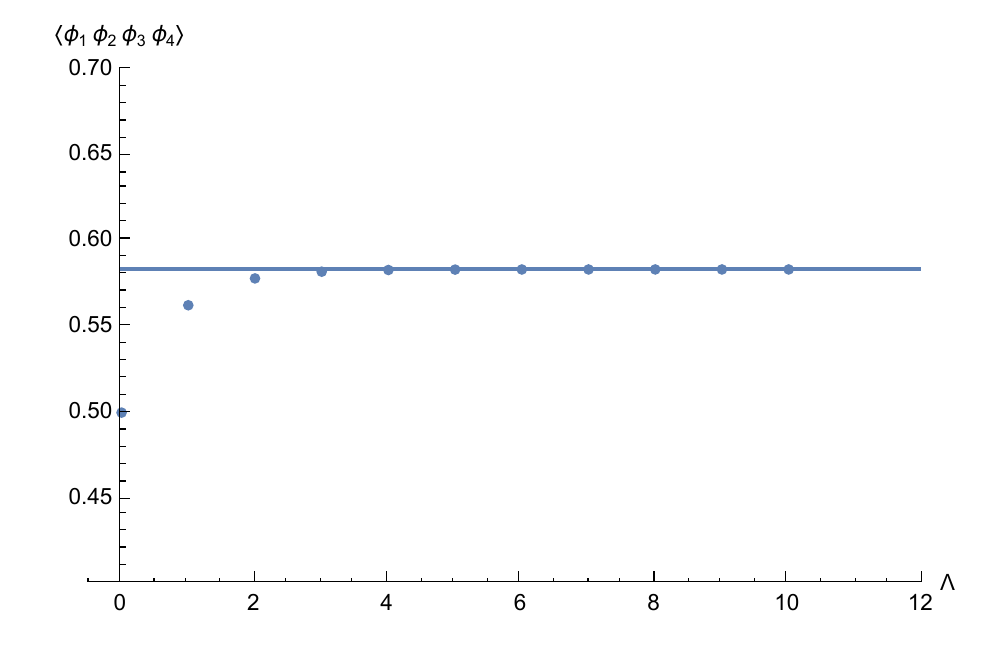}%
\caption{We compare the exact four-point function $\langle \phi_{1} \phi_{2} \phi_{3} \phi_{4} \rangle $ value with the corresponding series expansions obtained from the OPE. The exact value of the four point function is represented by a horizontal line at $7/12$. The series expansion is truncated with cut off $\Lambda$. The horizontal axis shows the value of $\Lambda$. Convergence is extremely rapid.}
\label{ConvergenceCheck}
\end{center}
\end{figure}

As an explicit check of OPE convergence, we calculate the s-channel OPE for the case $(x_1, x_2,x_3,x_4) = (1,2,3,4)$. This configuration was discussed in Section \ref{OPEConv}, where we concluded that according to \cite{Qiao:2020bcs} the $s$-channel OPE should converge. Our conventions are spelled out in the two point function
\bea
\langle\phi^a(x^+,x^-_1,x_1)\phi^a(x^+,x^-_2,x_2)\rangle={1\over |x_1-x_2|}
\eea
The exact value of the four-point function we study is
\bea
	\langle \phi^{a}(x^{+},x^{-}_{1},x_{1})  \phi^{a}(x^{+},x^{-}_{2} x_{2}) \phi^{b}(x^{+},x^{-}_{3},x_{3}) \phi^{b}(x^{+},x^{-}_{4},x_{4}) &=& \frac{1}{|x_{1}-x_{3}|} \frac{1}{|x_{2}-x_{4}|} + \frac{1}{|x_{1}-x_{4}|} \frac{1}{|x_{2}-x_{3}|}\nonumber\\
	&=& \frac{7}{12}.
\eea
To check convergence of the OPE, we now use formula (A.6) for each bilocal. This corresponds to the $s$-channel OPE. Defining
\bea
	x &=& \frac{1}{2}(x_{1}+x_{2}), \hspace{20pt} y = \frac{1}{2}(x_{1} - x_{2})\\
	z &=&\frac{1}{2}(x_{3}+x_{4}), \hspace{20pt} w = \frac{1}{2}(x_{3} - x_{4}),
\eea
we have 
\bea
	&&\langle \phi^a(x^+,x^-_1,x_1)  \phi^a(x^+,x^-_2 x_2) \phi^b(x^+,x^-_3,x_3) \phi^b(x^+,x^-_4,x_4)\rangle =\nonumber\\
	&& \hspace{50pt} \sum\limits^{\infty}_{s,s'=0} \sum\limits^{\infty}_{d,d'=0} c_{sd}c_{s'd'} \left( y \partial_{x} \right)^{2d} \left( w \partial_{z} \right)^{2d'} \langle J_{2s}(y,x) J_{2s'}(w,z) \rangle .
\eea
Each term in the sum on the RHS has a definite conformal dimension given by $2s+2s'+2d+2d'+1$. We truncate the sum with a cut on the dimension of the operators summed, implemented as $2s+2s'+2d+2d'\le\Lambda$. We then compare the truncated sum to the exact result given by $7/12$. The results, up to $\Lambda=10$ are shown in Figure \ref{ConvergenceCheck}. The numerical results are convincing evidence indicating that the OPE convergences.

The configuration $(x_1,x_2,x_3,x_4)=(1,3,2,4)$ was also discussed in Section \ref{OPEConv}. According to \cite{Qiao:2020bcs} the $s$-channel OPE should not converge. For this configuration, the exact value of the four point function is $4/3$. Again truncating the series obtained from the OPE, with cut off values $\Lambda = 0,1,2, \cdots , 7$ we obtain
\bea
	\{2, 6, 22, 86, 342, 1366, 5462, 21846\}
\eea
for the value of the sum. The sum is clearly diverging.

\begin{figure}[h]%
\begin{center}
\includegraphics[width=0.65\columnwidth]{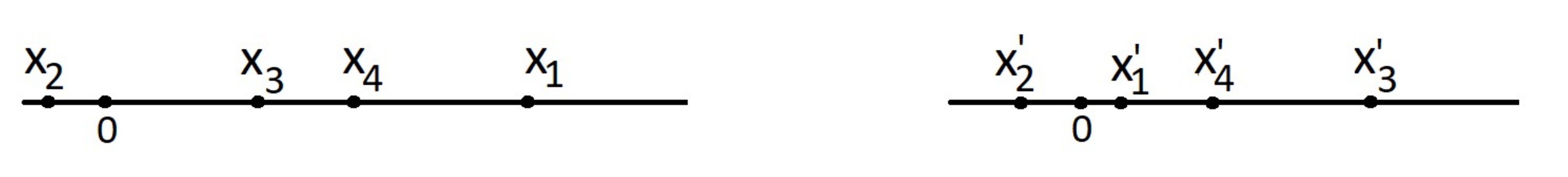}%
\caption{The initial configuration, shown on the left, is of type (B) in Figure \ref{fig:OPEconvConfg}. Perform an inversion about a point located between $x_2$ and $x_3$. This leaves point $x_2$ as the left most point, but reverses the order of $x_1,x_3,x_4$ so that we land up in a configuration of type (A) in Figure \ref{fig:OPEconvConfg}. The final configuration is shown on the right.}
\label{InvAct}
\end{center}
\end{figure}

Finally, the last configuration we study is $(x_1,x_2,x_3,x_4) = (6,-1,3,4)$. This configuration was also discussed in Section \ref{OPEConv}. It corresponds to a configuration of type (B) in Figure \ref{fig:OPEconvConfg} and according to \cite{Qiao:2020bcs} the $s$-channel OPE converges. Implementing a simple application of the midpoint OPE rule (\ref{explicitOPE}), we numerically find that the OPE does not converge. This is also the case if we use the OPE (\ref{SecondExplicitOPE}). To find the convergent OPE expansion it is useful to transform to a different conformal frame. As we have already seen, the four point function (and the conformal cross ratios) depend only on the coordinate transverse to the light cone. To move to the new conformal frame, we start with a translation (if needed) to position the origin between points $x_2$ and $x_3$ as shown in Figure \ref{InvAct}. We then apply the conformal inversion operation which takes
\bea
I\,:\, x^\mu\to x^{\prime\mu}={x^\mu\over x\cdot x}
\eea

We have the equality
\bea
&&\langle \phi^a(x^+,x^-_1,x_1)  \phi^a(x^+,x^-_2 x_2) \phi^b(x^+,x^-_3,x_3) \phi^b(x^+,x^-_4,x_4)\rangle\cr\cr
&&\qquad=\,\,
\langle I\cdot I \phi^a(x^+,x^-_1,x_1)I\cdot I \phi^a(x^+,x^-_2 x_2) 
I\cdot I\phi^b(x^+,x^-_3,x_3)I\cdot I \phi^b(x^+,x^-_4,x_4)I\cdot I\rangle
\nonumber
\eea
which is true since inversion squares to the identity $I\cdot I=1$. Since the free scalar field has $\Delta={1\over 2}$ we know that
\bea
I \phi^a (x_i^\mu) I = (x_i'\cdot x_i')^\Delta \phi'{}^a(x_i^{\prime\mu})
= \sqrt{x_i'\cdot x_i'}\phi'{}^a(x_i^{\prime\mu})\qquad i=1,2
\eea
Assuming that the conformal field theory vacuum is invariant under $I$ we now easily find
\bea
&&\langle \phi^a(x^+,x^-_1,x_1)  \phi^a(x^+,x^-_2 x_2) \phi^b(x^+,x^-_3,x_3) \phi^b(x^+,x^-_4,x_4)\rangle\cr\cr
&&\qquad=\,\,\sqrt{\prod_{i=1}^4 x_i'\cdot x_i'}
\langle \phi^{\prime a}(x^{\prime +},x^{\prime -}_1,x_1')
\phi^{\prime a}(x^{\prime +},x^{\prime -}_2 x_2') 
\phi^{\prime b}(x^{\prime +},x^{\prime -}_3,x_3')
\phi^{\prime b}(x^{\prime +},x^{\prime -}_4,x_4')\rangle
\nonumber
\eea
We are now in a configuration of type (A) so that we can apply the mid point OPE given in (\ref{explicitOPE}) to obtain a convergent $s$-channel OPE.

\subsection{Converting between $\eta$ and $\sigma$}

Both $\eta(x^+,x_1^-,x_1,x_2^-,x_2)$ and $\sigma(x^+,x_1^-,x_1,x_2^-,x_2)$ appear. They are defined as
\bea
\sigma(x^+,x_1^-,x_1,x_2^-,x_2)=\phi^a(x^+,x_1^-,x_1)\phi^a(x^+,x_2^-,x_2)
\eea
and
\bea
\eta(x^+,x_1^-,x_1,x_2^-,x_2)=:\phi^a(x^+,x_1^-,x_1)\phi^a(x^+,x_2^-,x_2):
\eea
so that we can write the operator equation
\bea
\sigma(x^+,x_1^-,x_1,x_2^-,x_2)=\eta(x^+,x_1^-,x_1,x_2^-,x_2)
+\langle \sigma(x^+,x_1^-,x_1,x_2^-,x_2)\rangle
\eea
This is an operator equation so it can be used inside any correlation function. As an example
\bea
\langle\eta(x^+,x_1^-,x_1,x_2^-,x_2)\eta(x^+,x_3^-,x_3,x_4^-,x_4)\rangle
&=&\langle\sigma(x^+,x_1^-,x_1,x_2^-,x_2)\sigma(x^+,x_3^-,x_3,x_4^-,x_4)\rangle\cr\cr
&-&\langle\sigma(x^+,x_1^-,x_1,x_2^-,x_2)\rangle
\langle\sigma(x^+,x_3^-,x_3,x_4^-,x_4)\rangle\cr\cr &&
\eea
Consequently any expectation value of $\eta$'s can be turned into an expectation values involving only $\sigma$'s.

\end{appendix}

\end{document}